\documentclass[3p, review,number,sort&compress]{elsarticle}

\begin{document}

\begin{frontmatter}

\title{Development of an intense positron source using a crystal--amorphous hybrid target for linear colliders}
\author[huniv]{Y.~Uesugi\corref{cor1}}
\author[huniv]{T.~Akagi}
\author[ipnl]{R.~Chehab}
\author[lal]{O.~Dadoun}
\author[keka]{K.~Furukawa}
\author[keka]{T.~Kamitani}
\author[huniv]{S.~Kawada}
\author[kekp]{T.~Omori}
\author[huniv]{T.~Takahashi}
\author[keka]{K.~Umemori}
\author[keka]{J.~Urakawa}
\author[keka]{M.~Satoh}
\author[]{V.~Strakhovenko\corref{cor2}}
\author[keka]{T.~Suwada}
\author[lal]{A.~Variola}
\address[huniv]{Graduate School of Advanced Sciences of Matter, Hiroshima University, 1-3-1 Kagamiyama, Higashi-Hiroshima, Hiroshima 739-8530, Japan}
\address[keka]{Accelerator Laboratory, High Energy Accelerator Research Organization (KEK), 1-1 Oho, Tsukuba, Ibaraki 305-0801 Japan}
\address[kekp]{Institute of Particle and Nuclear Studies, High Energy Accelerator Research Organization (KEK), 1-1 Oho, Tsukuba, Ibaraki 305-0801 Japan}
\address[lal]{Laboratoire de l'acc\'{e}l\'{e}rateur lin\'{e}aire (LAL), Universit\'{e} Paris--Sud 11, Batiment 200, 91898 Orsay Cedex, France}
\address[ipnl]{Institut de Physique Nucl\'{e}aire de Lyon (IPNL), Universit\'{e} Lyon 1, CNRS/IN2P3, Villeurbanne, France}
\cortext[cor1]{Corresponding author.
\textit{E-mail address}: uesg@huhep.org (Y. Uesugi).}
\cortext[cor2]{The deceased.}

	\begin{abstract}

In a conventional positron source driven by a few GeV electron beam, a high amount of heat is loaded into a positron converter target to generate intense positrons required by linear colliders, and which would eventually damage the converter target.
A hybrid target, composed of a single crystal target as a radiator of intense gamma--rays, and an amorphous converter target placed downstream of the crystal, was proposed as a scheme which could overcome the problem.
This paper describes the development of an intense positron source with the hybrid target.
A series of experiments on positron generation with the hybrid target has been carried out with a 8--GeV electron beam at the KEKB linac.
We observed that positron yield from the hybrid target increased when the incident electron beam was aligned to the crystal axis and exceeded the one from the conventional target with the converter target of the same thickness, when its thickness is less than about 2 radiation length.
The measurements in the temperature rise of the amorphous converter target was successfully carried out by use of thermocouples.
These results lead to establishment to the evaluation of the hybrid target as an intense positron source.

	\end{abstract}

	\begin{keyword}
Positron source \sep Crystal channeling \sep Electron and positron linear accelerators

	\end{keyword}

\end{frontmatter}


\pagestyle{plain}
\setcounter{page}{1}

	\section{Introduction}

The next--generation linear colliders, such as the International Linear Collider (ILC) \cite{ilc_interim_report_2011-lores} and the Compact Linear Collider (CLIC) \cite{CERN-OPEN-2008-021}, require a large amount of positrons in short time periods.
In particular, the ILC requires a positron beam with the bunch population of $2 \times 10^{10}$,  the bunch separation of 332 ns and 1,312 bunches in a bunch train with the train repetition rate of 5 Hz.
The design luminosity of the order of $10^{34}$cm$^{-2}$s$^{-1}$ equals the value of the KEKB synchrotron \cite{kekbsync}, which has the highest luminosity in electron--positron colliders.
To realize such high luminosity, the ILC requires acceleration and collision of intense electron and positron beams. 
The detail of the issues and the RD of the ILC is described in \cite{ilc_interim_report_2011-lores}. 
Among those issues, the development of an intense positron source is a key to realize the linear collider.

Conventionally, positrons are generated by impinging an electron beam of a few GeV into a converter target via electromagnetic cascade processes.
The converter target is typically a high--Z material such as the tungsten with a thickness of 4 to 6 radiation lengths, which is nearly equal to the depth of the shower maximum.
Since the converter target is easily damaged by the heat load associated with the development the cascade, conventional systems suffer from limitations on the positron yield due to the heat load.
The damage of the converter target was quantified under the operating condition of SLC at SLAC National Accelerator Laboratory (formerly Stanford Linear Accelerator Center) \cite{slac-r-571}.

A different approach for generating positrons, which would overcome the heat load problem, is impinging gamma--rays into the converter target instead of electrons.
By using combination of initial photons above O(10 MeV) and a thin high--Z material converter target, positrons are produced in the converter target dominantly via electron--positron pair production from initial photons and the energy deposition is reduced.
In addition, by keeping the distance between the source of photons and the converter target, the photon beam spot size on the converter target can be increased for relaxing the localized energy deposition.
If intensity of the incident gamma--ray is enough and the thickness of the converter target is appropriately chosen, it may be possible to reduce the heat load in the converter target while ensuring the required positron yield.
For example, the ILC has adopted the positron source driven by gamma--rays using a long magnetic undulator.

There is a method based on gamma--ray which uses a crystal assisted radiation.
It was proposed by R. Chehab and his colleagues at LAL d'Orsay \cite{PAC1989_0283}, and experimental studies has been carried out with the 1.3--GeV synchrotron in Institute for Nuclear Study, the University of Tokyo \cite{PhysRevLett_80_1437}, at CERN with the experiment WA 103 \cite{cern1, cern2} and the KEKB linac at KEK \cite{PhysRevE.67.016502}.
Simulations have shown \cite{091003} that a crystal converter presents the advantage of less energy deposition than the equivalent amorphous converter (giving the same positron yield).
However, to avoid significant energy deposition in the crystal, due to large incident beam power, which could affect the available string potential -- due to thermal vibrations -- a separation between the thin crystal--radiator and the thicker amorphous converter led to the hybrid source scheme.
Moreover, using a dipole magnet in between to remove the charged particles is reducing considerably the energy deposition and the peak energy deposition density.
This last scheme is considered, here, as the hybrid target.
In 2007, the hybrid target scheme was proposed by Chehab, Strakhovenko and Variola \cite{10} as a possible way to generate intense positrons required by CLIC or ILC.
In this article, we report results of experimental studies to investigate feasibility of the hybrid target as an intense positron sources.

\clearpage
	\section{A crystal assisted radiation}

\begin{figure}[h]
\centering
\includegraphics[width=.5\textwidth]{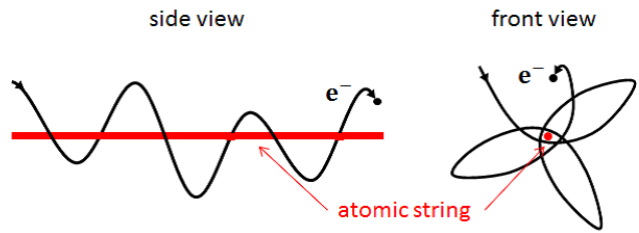}
\caption{Drawing of the electron's trajectory in axial channeling.}
\label{screw_trajectory}
\end{figure}

When a charged particle enters parallel to a crystal axis or planes in a crystal material, the transverse electric field of the axis or the planes constrains the particle motion in the transverse direction; it is known as the channeling effect.
In particular, an electron is captured by the strong field of an atomic string, which is called the axial channeling \cite{chr}.
The average crystalline potential of an isolated atomic string $U(r)$ is approximately expressed as
\begin{equation}
U(r)	=	-\frac{Z e^{2}}{d}\ln\left( 1 + \frac{3 a^{2}_{TF}}{r^{2}} \right),
\label{lin}
\end{equation}
where $a_{TF}$ is the Thomas-Fermi screening radius, $Z$ is the atomic number of the crystal, $d$ is the interatomic spacing in the string, $e$ is the electric charge, and $r$ is the distance from the axis \cite{Lindhard}.
If the transverse kinetic energy $p^{2}_{T}c^{2}/2E$ of the incident electron with respect to the strings is smaller than the depth $U_{0}=\left|U(0)\right|$ of the potential and if the entrance point is close enough to a string, the electron is trapped in the potential well and travels along the axis for some distance depending mainly on its energy.
The transverse energy of the electron can be written in the form
\begin{equation}
E_{T} = \frac{p^{2}_{T}c^{2}}{2E} + U(r),
\label{eq1}
\end{equation}
where $p_{T}$ is the transverse component of the electron momentum $p$, and $c$ is the velocity of light.
Considering that $p_{T}$ is much smaller than $p$, the incident angle $\psi$ with respect to the axis can be written as $\psi=p_{T}/p_{z}$, and we can transpose expression (\ref{eq1}) to
\begin{equation}
E_{T} = \frac{pv}{2}\psi^{2} + U(r),
\label{eq2}
\end{equation}
where $v$ is the velocity of the electron.
The channeled electron, which is in a bound state of $E_{T} < 0$ moves along a screw--like trajectory about the axial direction as shown in a drawing of figure \ref{screw_trajectory}.
The incident electrons have a critical angle of capture $\psi_{L}$, so--colled the Lindhard angle:
\begin{equation}
\psi_{L} = \sqrt{\frac{2U_{0}}{pv}}.
\label{eq3}
\end{equation}
For the crystal axis $\langle 111 \rangle$ of tungsten used in this experiment, the $U_{0}$ is 979 eV, and the angle $\psi_{L}$ is 0.495 mrad at an incident electron energy of 8 GeV.

\begin{figure}[h]
\centering
\includegraphics[width=.5\textwidth]{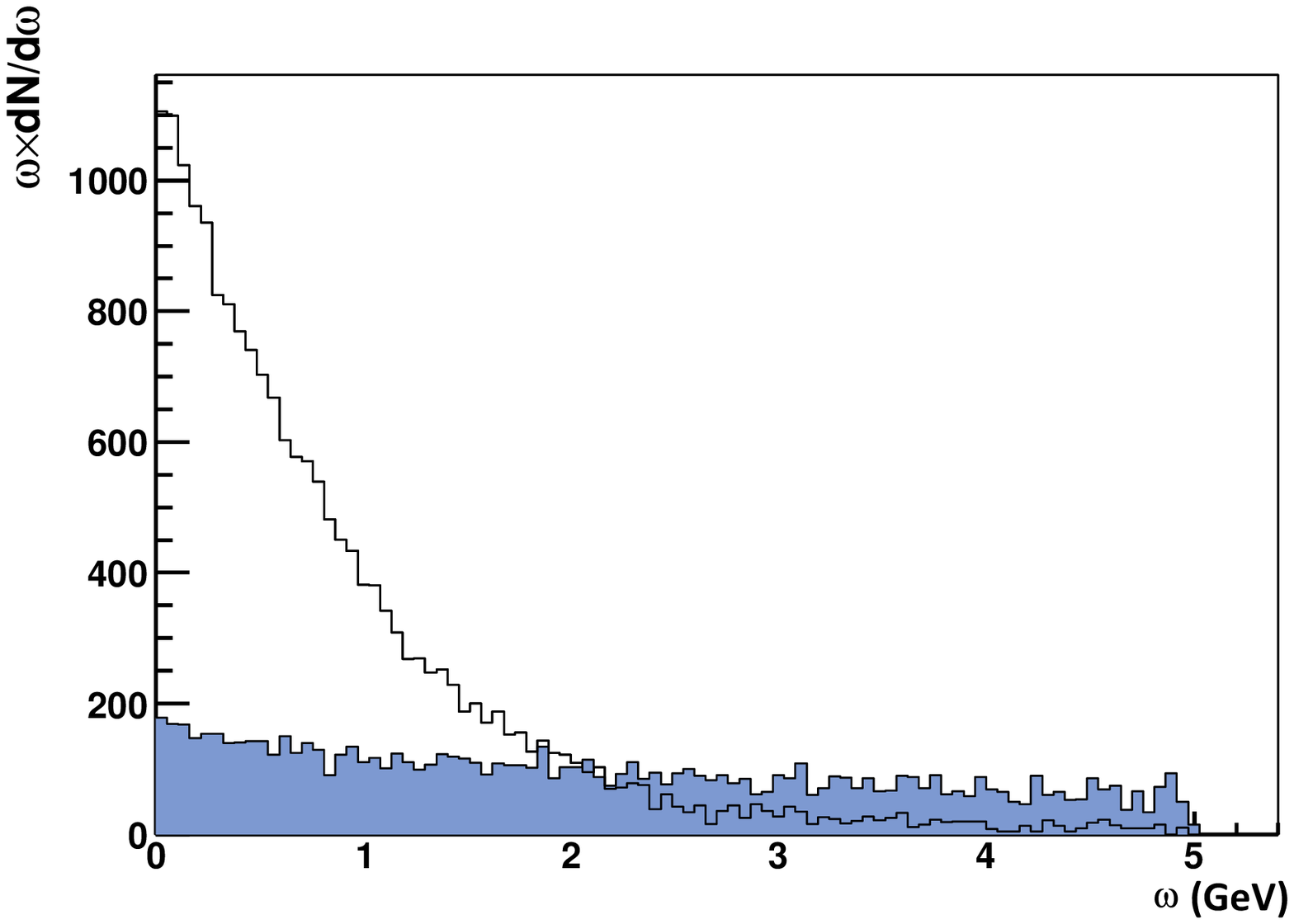}
\caption{The comparison of the emerging photon spectra between a pure bremsstrahlung (filling check) and from a channeling radiation (filling white), from a 5--GeV electron impinging on a 1.4 mm of tungsten crystal and amorphous target.}
\label{brems}
\end{figure}

Channeled electrons emit strong radiation; the \textit{channeling radiation}.
The intensity of the radiation is expressed in the form
\begin{equation}
I=\frac{2e^{2}}{3c^{3}}\dot{v_{T}}^{2}\gamma^{4},
\label{eq4}
\end{equation}
where $\dot{v_{T}}$ is the transverse acceleration of the electron which moves according to the condition of (\ref{eq2}), and  $\gamma=1/\sqrt{1-v^{2}/c^{2}}$ \cite{1-s2.0-0375960176904382-main}.
The channeling radiation has much larger intensity and shorter wavelengths than that of a magnetic undulator, due to the strong field of atoms that causes a short period oscillation with the order of some $\mathrm{\mu m}$ in the longitudinal direction with the amplitude of the order of an interatomic distance.
For the $\langle 111 \rangle$ axis of tungsten with the incident electron energy of 8 GeV, it was found by numerical simulations that the intensity of low energy gamma--rays ($\leq$ 30 MeV) is ten times higher than that of nominal bremsstrahlung.
Figure \ref{brems} shows the comparison of the photon spectra obtained from a 1.4 mm of tungsten crystal and a 1.4 mm tungsten amorphous converter, the electron energy being 5 GeV.

When electrons enter into the crystal with larger angle than $\psi_{L}$, they are no longer trapped by the crystallline potential but are still affected by it.
This situation has been discussed in \cite{vm}: the radiation becomes synchrotron like, below a critical incidence angle $\psi_{0}=U_{0}/mc^{2}$, where $m$ is the rest mass of an electron.
The angle $\psi_{0}$ is 1.92 mrad for $U_{0}=979$ eV which is about 4 times larger than the critical angle of channeling effect.

There is another effect known as coherent bremsstrahlung, which is the coherent radiation of successive planes or strings interfere \cite{cbb, cbbb}. 
Coherent bremsstrahlung has the spectrum with high-order peaks caused by the condition of interference as,

\begin{equation}
\omega_{n}(\theta)	=	\frac{v\psi}{d(1-\beta \cos \theta)} 2 \pi n	\quad	(n=1, 2, 3...),
\label{cb}
\end{equation}
where  $\omega_{n}(\theta)$ is the circular frequency of the photon, $\theta$ is the photon emission angle, $\beta=v/c$, and $d$ is the distance between the atoms.
In the conditions of an ultrarelativistic case ($\beta \sim 1-1/2\gamma^{2}$) and forward directing photons ($\theta =0$), the expression (\ref{cb}) is approximated to 
\begin{equation}
\omega_{n}(0) = \frac{4\pi c \psi \gamma^{2}}{d}n,
\label{cb2}
\end{equation}
where $\gamma = (1-\beta^{2})^{-1/2}$ is relativistic gamma factor of the incident electron.
When the energy of incident electrons $E$ is 8 GeV, $\hbar\omega = 4$ GeV $(=E/2)$ for $\psi=2.08$ mrad.
Therefore the contribution from coherent bremsstrahlung could be expected up to the energy region around a few GeV in the experiment.

\clearpage
	\section{Experimental setup and schemes}

\begin{figure}[h]
\centering
\includegraphics[width=.5\textwidth]{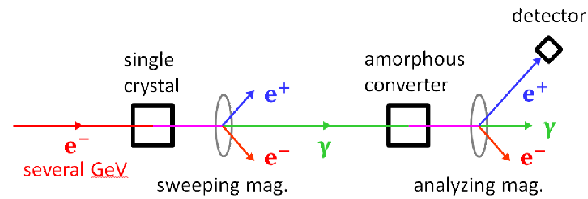}
\caption{Construction of the hybrid target using a single crystal target and an amorphous converter target.}
\label{hybrid_img}
\end{figure}

\begin{figure}[h]
\centering
\includegraphics[width=\textwidth]{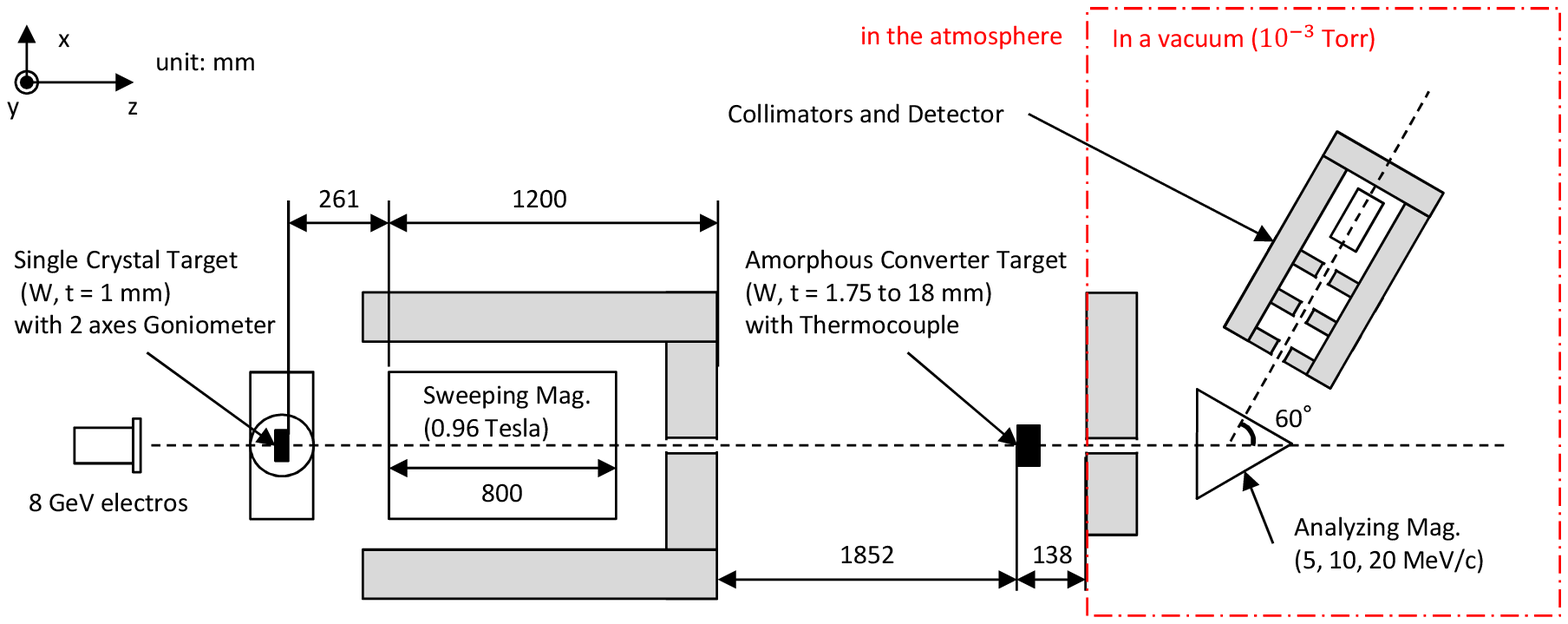}
\caption{Layout of the hybrid target and the positron detector system in the experiment in top view.}
\label{setup}
\end{figure}

A series of experiments on positron generation were performed at KEK, Tsukuba, Ibaraki, Japan.
A schematic drawing of the hybrid target, which was composed of a single crystal target to produce gamma--rays and an amorphous converter target made of tungsten to generate positrons, is shown in figure \ref{hybrid_img},
and the experimental setup installed in the 3rd switchyard of the KEKB linac \cite{kekb} is shown in figure \ref{setup}.
The KEKB linac delivers single--bunch electron beams with energy of 8 GeV.
The beam parameters are given in table \ref{para}.
The single crystal target made of tungsten was mounted on the 2--axes goniometer so as to align the crystal axis to the incident electron beam.
The thickness of the crystal target was 1 mm and $\langle 111 \rangle$ axis was aligned to the electron beam.
The surface mosaicity of this crystal on both sides was measured by an X-ray scattering method was $\sim 0.5$ mrad.
After the crystal target, a sweeping magnet with a length of 800 mm and a magnetic field of 0.96 tesla was installed to sweep out secondary charged particles generated in the crystal target.
Thus, only gamma-rays were directed to the amorphous converter target installed 3313 mm downstream of the crystal target.
Five amorphous converter targets of different thicknesses, 1.75, 3.5, 5.25, 8 and 18 mm, were mounted on a linear stage to place the converter target of desired thickness on the beam line during the experiment.
The particles from the converter target were directed into a vacuum chamber kept at pressure of $10^{-3}$ Torr.
Positrons with momenta of 5, 10 or 20 MeV/$c$ were selected by a magnetic momentum analyzer and led into the detector through collimators.
In between the analyzing magnet and the detector, 3 collimators of aperture of 20 mm were placed to reduce background as well as to improve the momentum resolution.
The geometrical acceptance of the detector was 0.22 msr which was defined by the collimator just before the detector.
For the data analysis, which is described in section 4. 2, the efficiency of the detector system was evaluated by the simulation which took into account geometrical configuration of all components.
The detector comprised lucite Cherenkov counter, its output is proportional to the number of incident positrons to the detector.
Thermocouples were stuck on the back surface of two amorphous converter targets (8 mm and 18 mm) to measure the temperature from which we expect to get information of the heat load on the converter target due to the cascade shower.

\begin{figure}[h]
\centering
\includegraphics[width=.5\textwidth]{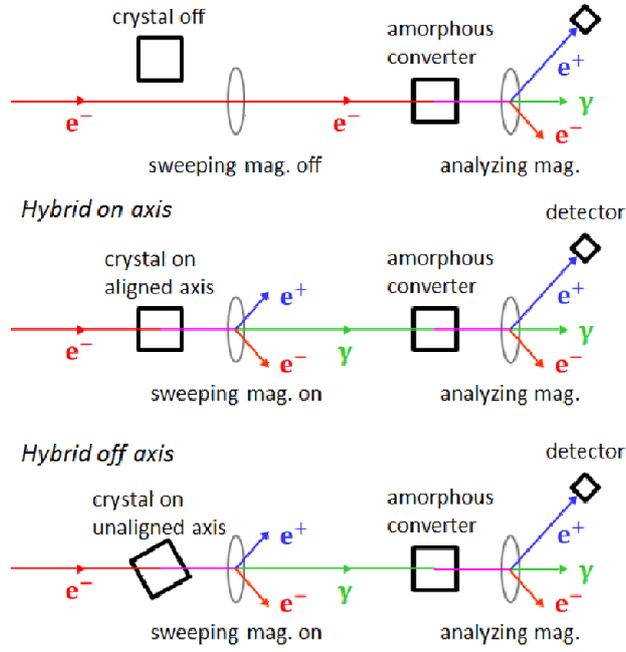}
\caption{Three schemes in the target configuration.}
\label{scheme}
\end{figure}

The positron yield was investigated in three experimental setup as shown in figure \ref{scheme}.
The first one was the \textit{conventional} scheme in which the incident electron beam with energy of 8 GeV directly irradiated the amorphous converter target.
The second was the \textit{hybrid on axis}, and the last is the \textit{hybrid off axis}.
In these schemes, the single crystal target is inserted into the beamline.
Charged particles coming out from the crystal were swept out by the sweeping magnet and only photons from the crystal were directed to the amorphous converter target.
The crystallographic axis is aligned with the direction of the incident electron beam at the \textit{on axis}, but not at the \textit{off axis} scheme, respectively.
Therefore, the enhancement effect by the crystal assisted radiation occurs only in second scheme. 
In these three schemes, five converter targets with different thicknesses were tested in order to measure the thickness dependence in the generated positron yield.
Momentum dependence in the positron yield was also measured with the analyzing magnet.
The contribution of the background for the \textit{conventional} scheme were estimated by the data taken without converter targets on the beamline.
As for the background estimation in the hybrid schemes, the crystal was removed from the beamline and the sweeping magnet was turned on; only the synchrotron radiation from incident electrons was irradiating the converter target.

Transverse sizes of the incident electron beam on the single crystal target and the amorphous converter target were measured by using a fluorescent screen before and after the experiment.
The standard deviation of the two measurements are shown in table \ref{spot}.
It is found that the profile of the electron beam was elliptic at the single crystal target but almost circular at the converter target.
The angular spread of the electron beam was estimated from the beam sizes at the position of the crystal target and the converter target, and were 0.15 mrad for horizontal (x) and 0.030 mrad for vertical (y) directions, respectively.

\begin{table}[h]
\begin{center}
\begin{tabular}{lll} \hline
Parameter 		&Value		&Unit	\\ \hline
Energy			&8	&GeV	\\
Repetition		&50 (max)	&Hz	\\
Charge in a pulse	&1 (typ), 3 (max)	&nC	\\ 
Bunch length		&10 (FWHM)	&ps	\\ \hline
\end{tabular}
\caption{Properties of the incident electron beam.}
\label{para}
\end{center}
\end{table}

\begin{table}[h]
\begin{center}
\begin{tabular}{lll} \hline
Target position		&$\sigma_{x}$ (mm)	&$\sigma_{y}$ (mm)	\\ \hline
Single crystal		&$0.89\pm0.01$		&$1.71\pm0.07$		\\
Amorphous converter	&$1.86\pm0.03$		&$1.92\pm0.06$		\\ \hline
\end{tabular}
\caption{Transverse spot sizes of the incident electron beam at the positions of the single crystal target and the amorphous converter target.}
\label{spot}
\end{center}
\end{table}

\clearpage
	\section{Measurement of the positron yield}

		\subsection{Goniometer angle dependence of the positron production}

\begin{figure}[h]
\centering
\includegraphics[width=.5\textwidth]{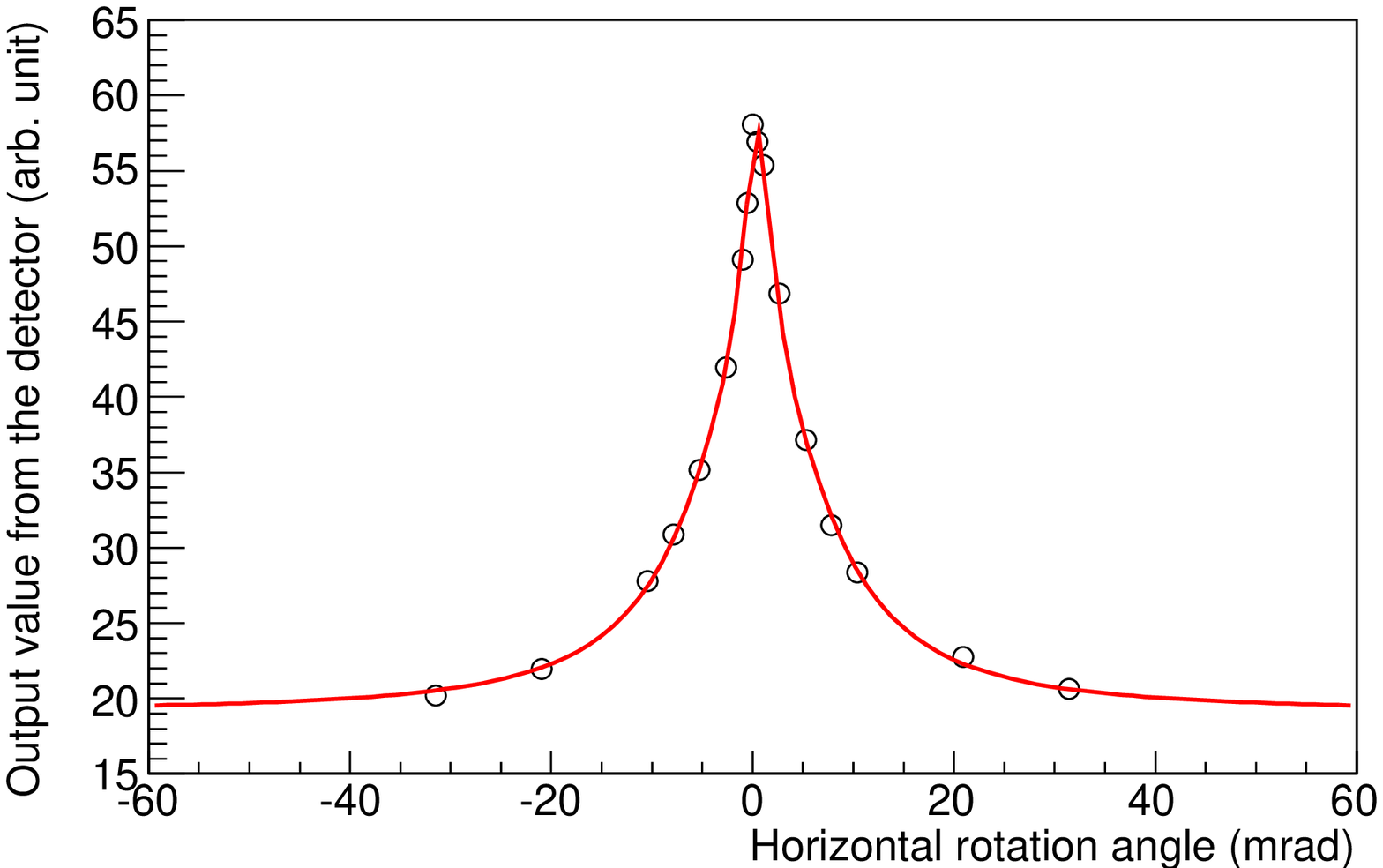}
\includegraphics[width=.5\textwidth]{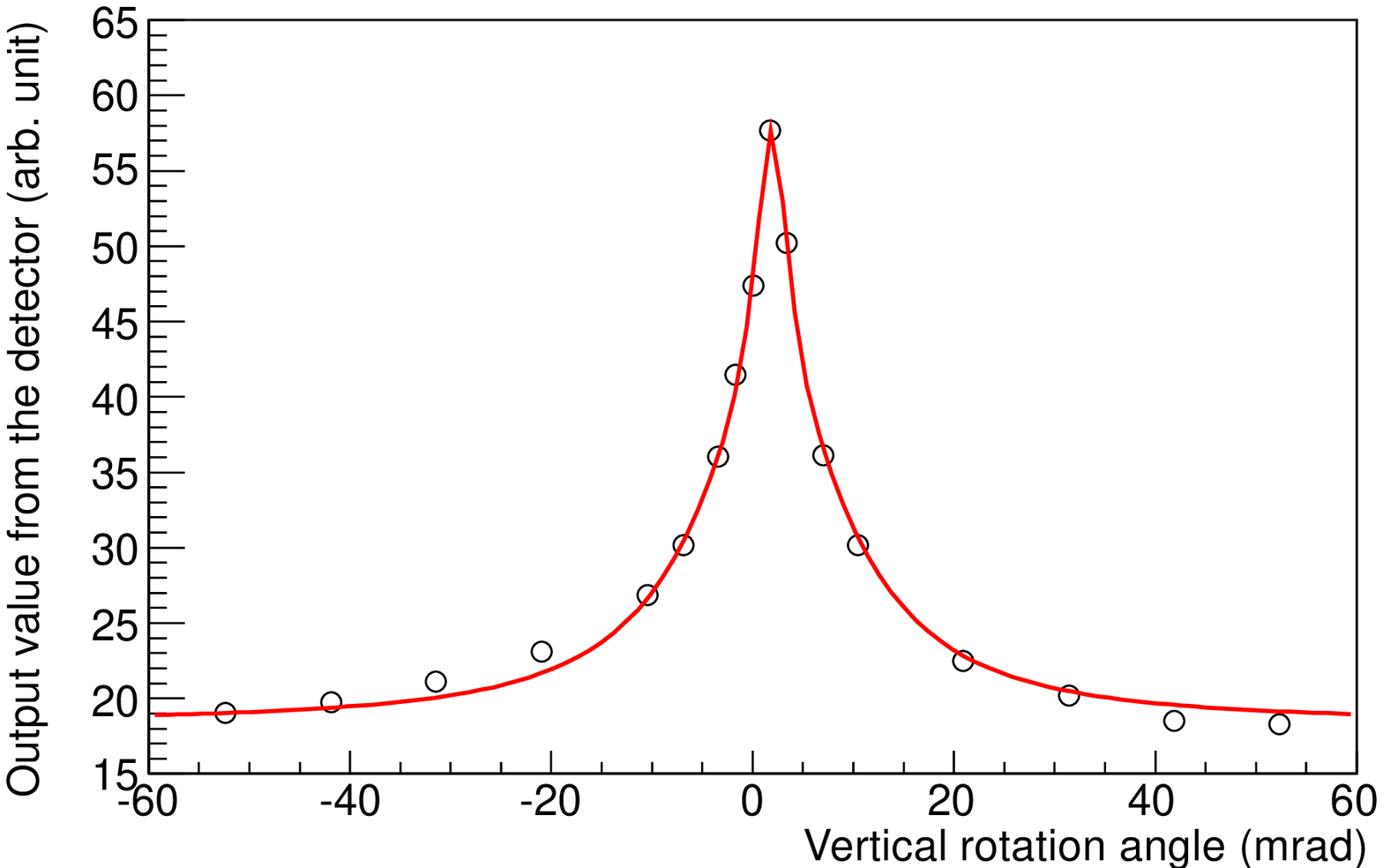}
\caption{Goniometer angle dependence of the positron production (\textit{rocking curve}). Top is the result with regards to horizontal rotation, and bottom is that to vertical rotation. When measuring one angle of rotation, the other was being kept equal to zero.}
\label{rcurve}
\end{figure}

To confirm the intensity enhancement of the gamma--rays by the crystal, the positron yield was measured while changing relative angle between the crystalline axis $\langle 111 \rangle$ and the electron beam axis.
The enhancement of the positron yield at the specific angle (\textit{rocking curve}) was observed as shown in figure \ref{rcurve}.
We empirically found that the data can be well fitted by a sum of two Lorentz functions:
\begin{eqnarray}
f(\theta)	&=&	A\frac{\Gamma_{1}}{(\theta-\langle\theta\rangle)^{2}+\Gamma^{2}_{1}}	\nonumber\\
		&\quad&	+\> B\frac{\Gamma_{2}}{(\theta-\langle\theta\rangle)^{2}+\Gamma^{2}_{2}}+\mathrm{const},
\label{rcurve_fit}
\end{eqnarray}
with $A$, $B$, $\langle\theta\rangle$, $\Gamma_{1}$ and $\Gamma_{2}$ being fitting parameters.
The results of the fitting are given in table \ref{gamma_angle}.
The facts that 1) the data were fitted with the sum of the two function and 2) the widths  $\Gamma_{1}$ and $\Gamma_{2}$ are wider than the critical angle of the channeling radiation (0.495 mrad) indicates the appreciable contribution  of several radiation processes described in section 2 and associated, in significant part, to above barrier particles.

\begin{table}[h]
\begin{center}
\begin{tabular}{lllllll} \hline
			&A		&B		&$\langle\theta\rangle$	&$\Gamma_{1}$ (mrad)	&$\Gamma_{2}$ (mrad)	&FWHM				\\ \hline
Horizontal		&$31\pm11$	&$179\pm11$	&$0.462\pm0.058$	&$1.78\pm0.36$		&$8.4\pm1.3$		&$8.7 ^{+2.5}_{-2.0}$	\\
Vertical		&$41\pm13$	&$196\pm16$	&$1.899\pm0.087$	&$2.05\pm0.37$		&$10.1\pm1.9$		&$8.9 ^{+3.3}_{-2.5}$	\\ \hline
\end{tabular}
\caption{The results of the fitting in the horizontal and vertical rocking curve.}
\label{gamma_angle}
\end{center}
\end{table}

\clearpage
		\subsection{Evaluation of the positron yield}

To evaluate the experimental data quantitatively, numerical simulations were perfomed by using the GEANT4 (9.4 Patch--01) \cite{geant4} and the FOT code \cite{1-s2.0-0168583X9090122B-main, fot}.
The FOT simulates electromagnetic interactions taking into acount channeling radiation, coherent and incoherent bremsstrahlung, and the transmission of electrons (or positrons) in the crystal.
We applied the FOT code to simulate the processes in the crystal for the case of the \textit{hybrid on axis} configuration.
The mosaic spread of the crystal was not taken into account in this simulation.
For the simulation of the \textit{hybrid off axis} configurations, we put amorphous tungsten of 1--mm thick and used the GEANT4 since processes in the \textit{off axis} case is same with the ones in the amorphous.
The positron intensities in the experimental results were normalized with those in the simulations obtained in \textit{conventional} scheme, because it is known that the GEANT4 is highly reliable in calculation accuracy and reproducibility.

\begin{figure}[h]
\centering
\includegraphics[width=.5\textwidth]{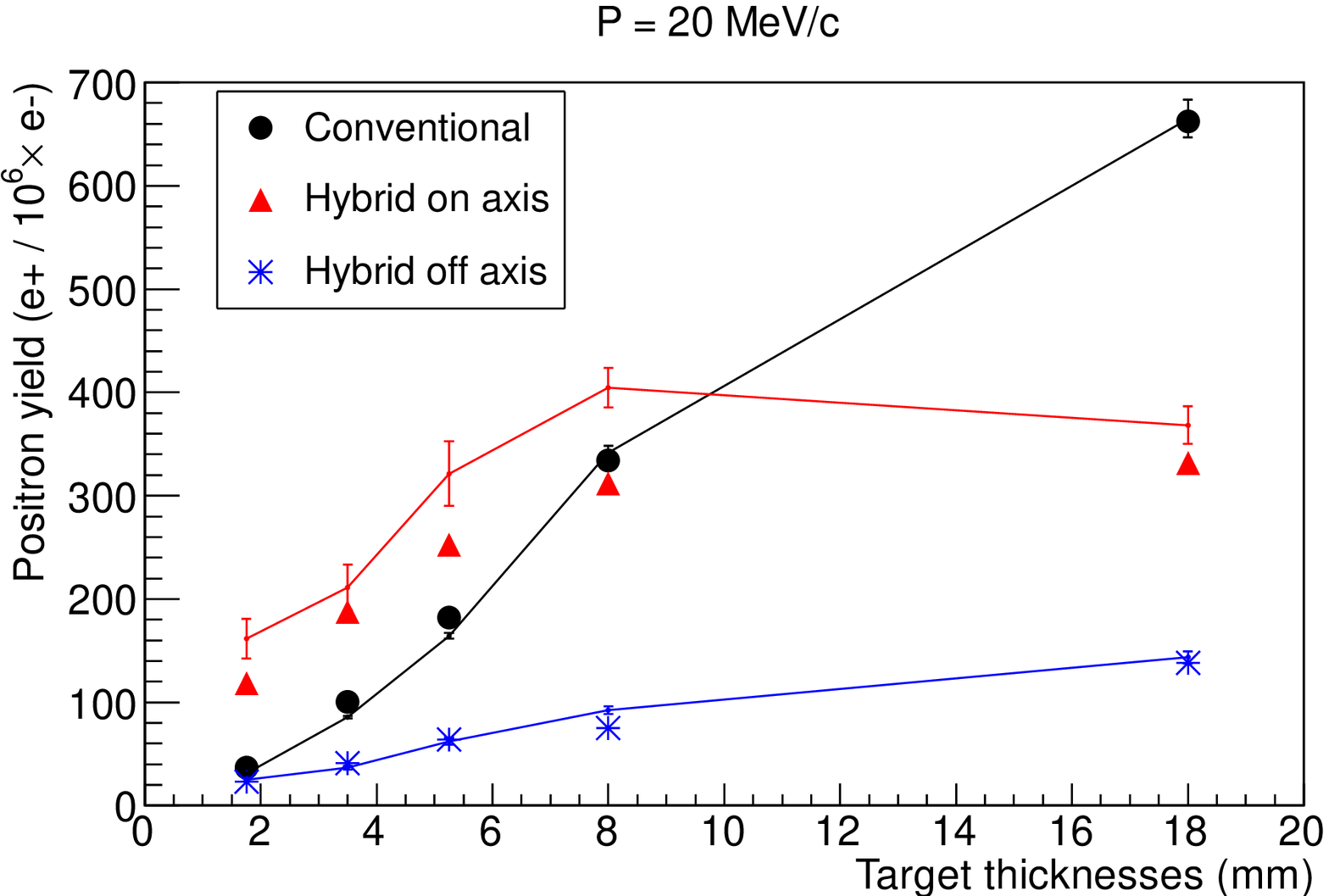}
\includegraphics[width=.5\textwidth]{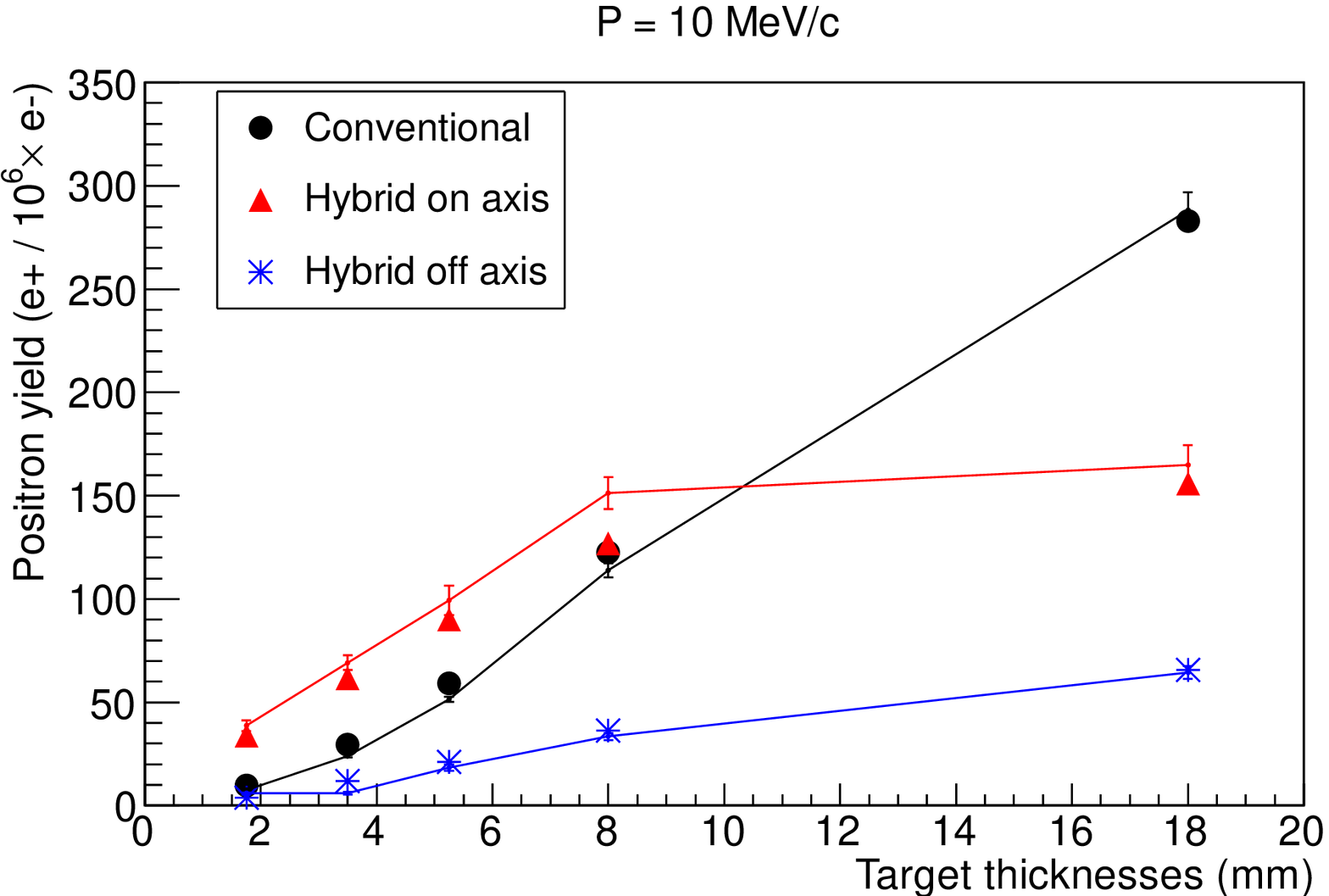}
\includegraphics[width=.5\textwidth]{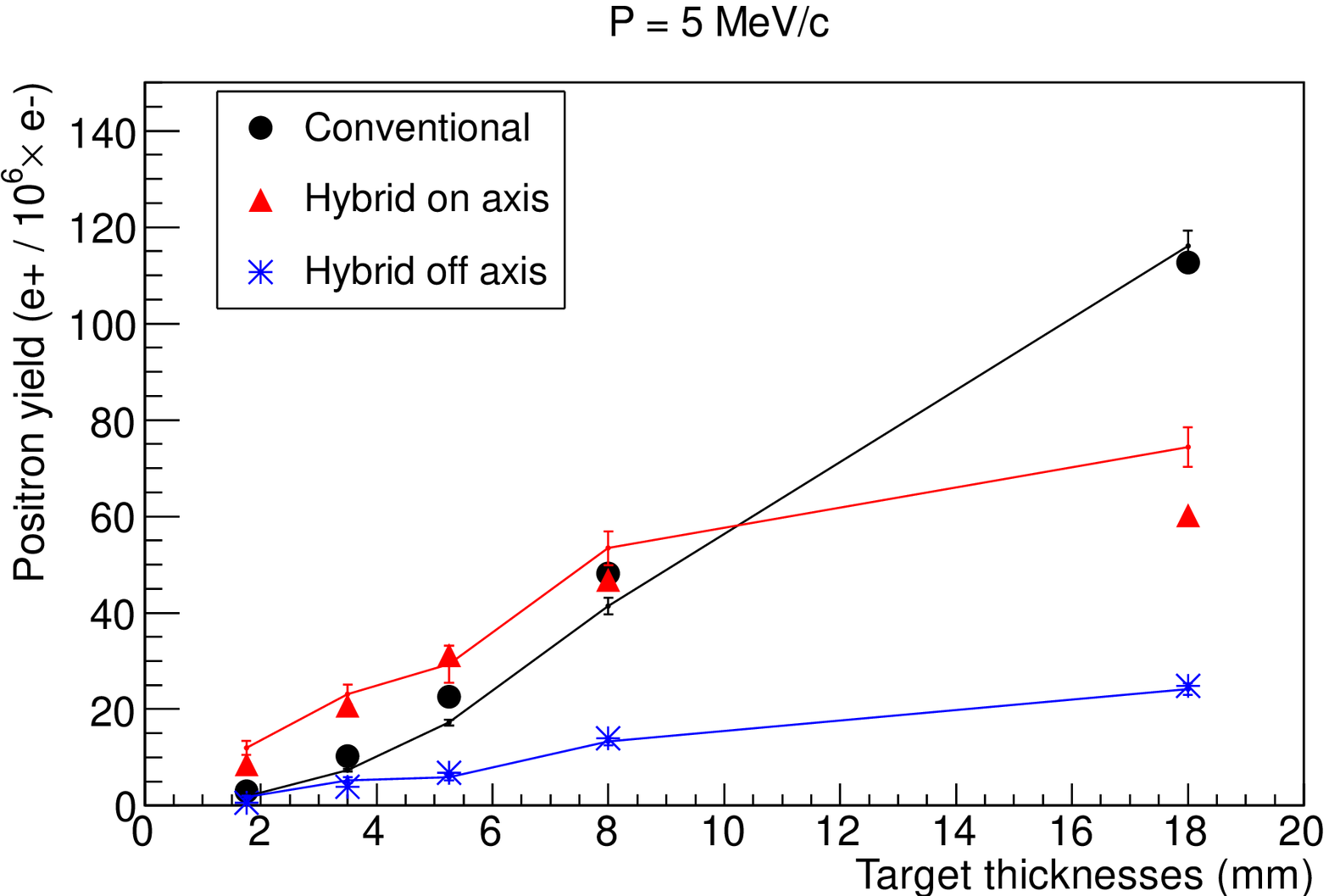}
\caption{Comparison of thickness dependences in the positron yield between experimental results and simulations in the analyzed momenta of 20 MeV/$c$, 10 MeV/$c$, and 5 MeV/$c$, from the top.
On these figures, markers show the experimental data, polygonal lines indicate the simulation, and the vertical axis means the positron yield per million electrons.}
\label{sep}
\end{figure}

Figure \ref{sep} shows the thickness dependence of the positron yield for the \textit{conventional}, the \textit{hybrid on axis} and the \textit{hybrid off axis} at the momentums of 20 MeV/$c$, 10 MeV/$c$ and 5 MeV/$c$, where the experimental data were normalized independently for each momentum.
In all analyzed momenta, the yields were the greatest at the thickness of 18 mm in the \textit{conventional} scheme.
This thickness equals to the the shower maximum of electromagnetic cascade for a 8--GeV electron beam and amorphous tungsten.
It should be mentioned that the positron yield in the \textit{hybrid on axis} scheme is greater than that of the \textit{conventional} scheme when the thickness of the converter target is thinner than 8 mm.
This result comes from a main characteristics of the hybrid target system; i.e. positrons can be directly created via pair creation.
In the case of the unaligned crystal axis, that is the \textit{hybrid off axis} scheme, the yield decreases against the \textit{on axis} as discussed in section 4.1.
Comparing the experimental data with the simulations, both results agree with each other on the thickness dependence in the \textit{off axis}.
As regards the \textit{on axis} case, the simulation is about 20 \% larger than the experiment while overall behavior is well reproduced.
It indicates that the evaluation of radiation processes in the crystal by FOT needs to be improved for the absolute yield of low energy photons.

\begin{figure}[h]
\centering
\includegraphics[width=.5\textwidth]{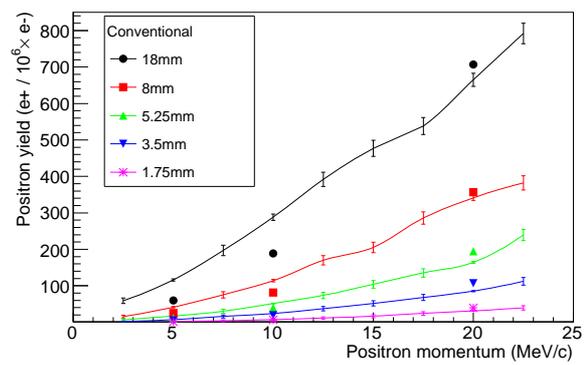}
\caption{Momentum dependences in the positron yield in the \textit{conventional} scheme. Here, one normalization constant is used, which is differing from the case of figure \ref{sep}.}
\label{mom_dist}
\end{figure}

The comparison in momentum dependence between experimental results and simulations is shown in figure \ref{mom_dist} for the \textit{conventional} scheme.
Unlike the comparison in figure \ref{sep}, a common normalization factor was used for all momenta.
If the acceptance were correctly reproduced, both should have agreed with each other regardless of the analyzed momenta.
It was unfortunately found that the simulation did not completely  reproduce the experiment in regard to momentum dependence.
This would mean that the experimental environment, e.g. the arrangement of collimators, was not reproduced accurately in the simulation.
However, it does not directly concern the estimate of FOT because of the high reliableness of GEANT4.
The simulation using GEANT4 is actually reproduced with high accuracy in the thickness dependence as shown in figure \ref{sep}.


\clearpage
	\section{Temperature measurement}

Estimation of the heat load in the converter target is an important information to evaluate survivability of the target in the positron source operation.
However, the direct measurement of the heat load caused by the electromagnetic shower is not a trivial task.
Therefore, simultaneously with the positron yield, we measured temperature of the converter target as an estimation of the heat load.

Type K thermocouples with a sensitive area of 1 mm $\times$ 1 mm were used to measure the temperature rise.
They were pasted on the back surface of the 8--mm thick and the 18--mm thick amorphous converter targets.
Compensating leadwires connected to the thermocouples were laid to the monitoring room.
The output from the thermocouple was calibrated with cold-junction compensation and amplified by a thermocouple converter.
The signal is recorded with a sampling rate of 1 kHz.
After adjusting the position of the thermocouple to the incident electron beam, the temperature rise, synchronized with electron bunches, were measured with the beam repetition rate of 1 Hz.
The converters are mounted on the linear stage with the aluminum holders so that the heat was evaluated to the environment (air) directly from the converter or via the holder.

\begin{figure}[h]
\centering
\includegraphics[width=.5\textwidth]{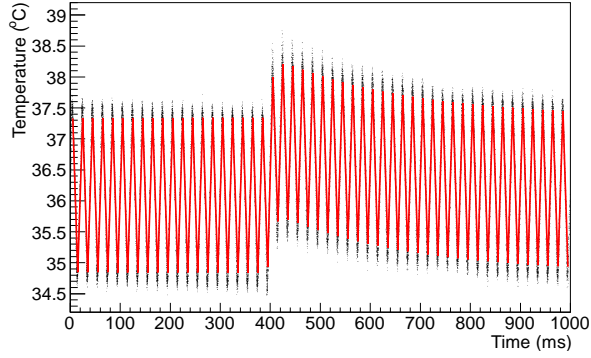}
\caption{Time trace in the temperature rise mesurement in the condition of the conventional, 18--mm thick amorphous converter target. The 50--Hz noise trace induced from power lines is clearly visible.}
\label{temp}
\end{figure}

Figure \ref{temp} shows the measured temperature rise in one second corresponding to an irradiation of single--bunch.
As is seen from the figure, there was a noise from the 50--Hz power line in the data.
To evaluate the temperature variations due to the electromagnetic shower in the converter, the data were fitted with the function $f(t)$ as
\begin{eqnarray}
f(t)	&=&	ae^{-(t+\delta+t_{0})/\tau_{1}} \left(1-e^{-(t+\delta+t_{0})/\tau_{2}}\right)	\nonumber	\\
	&\quad&	+\> b\sin\{50\cdot2\pi(t+\delta)\}				\nonumber	\\
	&\quad&	+\> \mathrm{const},
\label{fitf}
\end{eqnarray}
where $a, \tau_{1}, \tau_{2}, t_{0}, \delta, b$ are the amplitude of temperature rise, decay time and rise time of the temperature, time offset of the temperature rise, overall time offset including 50--Hz noise, and the amplitude of the noise, respectively.

\begin{figure}[h]
\centering
\includegraphics[width=.5\textwidth]{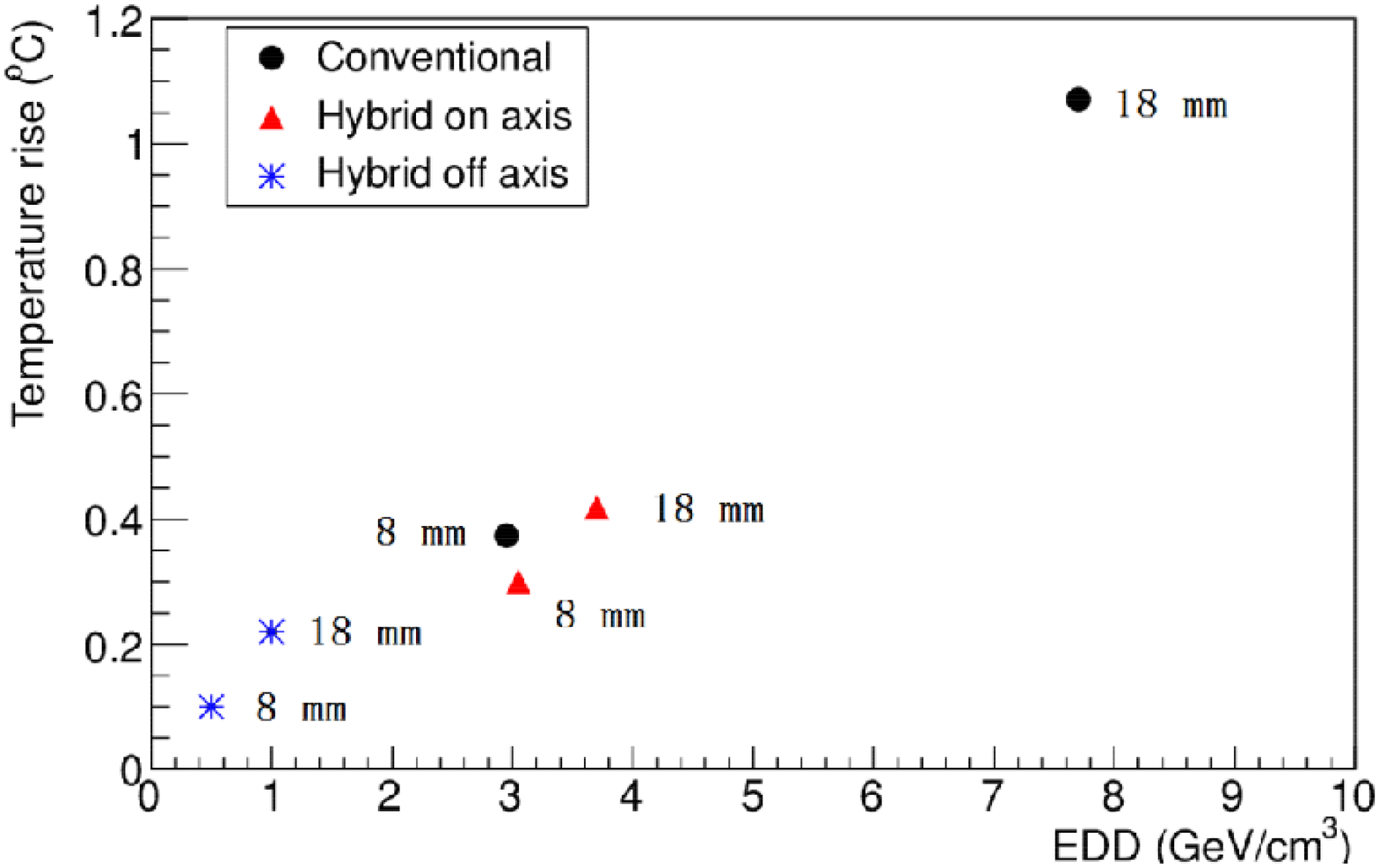}
\caption{Correlation plot between the measured amounts of the temperature rise and the calculation results in the EDD.}
\label{pedd}
\end{figure}

The result of the measurements in regard to three schemes of targets with the 8--mm thick and the 18--mm thick converter targets is shown in table \ref{temp_table}.
It was found that the slight temperature rise of 0.1 $^{\circ}$C, corresponding to the case of the \textit{hybrid off axis} scheme with the 8--mm thick, was able to be measured using the thermocouple.
The decay time constant was sufficiently greater than the resolution time of the measurement instrument.
Figure \ref{pedd} shows a correlation between the measured temperature rise and the energy deposition density (EDD) at the back end of converter targets computed with a simulation,
where the EDD is defined by unit of GeV/cm$^{3}$ and is the localized energy deposition in the converter target.
It is found that the temperature rise is proportional to the EDD and that the EDD could be evaluated by measuring the variation in the temperature rise measurements.

\begin{table}[h]
\begin{center}
\begin{tabular}{lllll} \\ \hline
Target scheme	& Thickness	& $a$: Temperature rise	& $\tau_{1}$: Decay time	& $\tau_{2}$: Rise time	\\
		& (mm)		& ($^{\circ}$C)		& (ms)				& (ms)			\\ \hline
Conventional	& 18		& $1.071\pm0.003$	& $332\pm3$			& $8\pm1$		\\
		& 8		& $0.373\pm0.003$	& $116\pm1$			& $3\pm1$		\\
Hybrid axis on	& 18		& $0.419\pm0.002$	& $537\pm1$			& $10\pm1$		\\
		& 8		& $0.300\pm0.002$	& $178\pm2$			& $2\pm1$		\\
Hybrid axis off	& 18		& $0.197\pm0.004$	& $542\pm3$			& $7\pm1$		\\
		& 8		& $0.095\pm0.001$	& $144\pm4$			& $3\pm1$		\\ \hline
\end{tabular}
\caption{Results of the temperature rise measurement. Each property corresponds to a parameter in the fitting function.}
\label{temp_table}
\end{center}
\end{table}

\clearpage
	\section{Conclusion}

We clearly observed the increase of positron yield when the crystal axis is aligned to the electron beam.
The width of the rocking curve was wider than the critical angle of the channeling effect or the angular spread of the incident electron beam, that was consistent with previous experiment indicating appreciable contribution of radiation from other than the channeling effect.
This fact is advantageous in practical operation, since it relaxes the precision of the alignment of the crystal target to the electron beam.

Since it has been confirmed that the simulation reproduces the experimental data, one would be able to perform the optimization in designing a positron source with the hybrid target by using the simulation.
According to the initial simulation, the angular spread of the positrons from the hybrid target was wider than the conventional target.
Because of small angular acceptance of current experimental set up, what we observed in figure \ref{sep} is smaller a part of the yields for the hybrid target than those for the conventional target.
In addition, the tungsten crystal may not be the best choice for the electron--photon converter.
Constructing the optimum target and positron capture section by using the simulation may give us the final conclusion of the ability to produce more positrons.

The result of temperature rise measurement gives us basic information about to evaluate the energy deposition in the positron converter target.
We need further study to evaluate relation between the measured temperature and energy deposition, however, it will be applicable to any positron target system once it is established.

To evaluate the yield performance in the hybrid target, a simulation including a capture system is planned.
To advance designs of the hybrid target, it is important to select the crystal material that gives the highest enhancement in gamma--rays, and to develop the converters with an efficient cooling system.
On the temperature measurement, we plan to measure a spatial distributions in the temperature rise measurement with arrayed thermocouples.

\clearpage
	\section*{Acknowledgment}

Authors would like to thank operation crew of the KEKB LINAC for providing excellent beams during the experiment.
This work was supported in part by JSPS KAKENHI Grant Number 2540314.

			\end{document}